\documentclass[preprint,singlecolumn,superscriptaddress,prl,floatfix,british]{revtex4}

\usepackage{amssymb}
\usepackage{amsmath}
\usepackage{babel}
\usepackage[sort&compress]{natbib}
\usepackage{topcapt}

\usepackage{graphicx}% Include figure files

\usepackage[colorlinks,linkcolor=dred,urlcolor=dblue,citecolor=dgreen]{hyperref}
\usepackage{stmaryrd}
\usepackage{amsfonts}
\usepackage{mathrsfs}
\usepackage{txfonts}
\graphicspath{{figs/}}

\usepackage{color}
\usepackage{pdfpages}
\definecolor{dred}{rgb}{.8,0.2,.2}
\definecolor{ddred}{rgb}{.8,0.5,.5}
\definecolor{dblue}{rgb}{.2,0.2,.8}
\definecolor{dgreen}{rgb}{.2,0.5,.2}

\usepackage{multirow}%talbe multirows

\newcommand{\Ref}[1]{(\ref{#1})}
\newcommand{\bra}[1]{\langle {#1}|}
\newcommand{\ket}[1]{| #1 \rangle}

\newcommand{\C}{\mathbb{C}}

\def\be{\begin{eqnarray}}
\def\ee{\end{eqnarray}}

\newcommand{\ca}{\mathcal A}

\newcommand{\ch}{\mathcal H}

\newcommand{\cs}{\mathcal S}

\renewcommand{\a}{\alpha}

\newcommand{\tr}{\mathrm{tr}}

\newcommand{\Ar}{\mathbf{Ar}}

\newcommand{\scap}[1]{\textsc{#1}}
\newcommand{\adscft}{\textsc{a}d\textsc{s}/\textsc{cft} }

\usepackage{times}

\begin{document}

\title{Measuring Holographic Entanglement Entropy on a Quantum Simulator}

\author{Keren Li}
\thanks{These authors contributed equally to this work.}
\affiliation{Shenzhen Institute for Quantum Science and Engineering, and Department of Physics, Southern University of Science and Technology, Shenzhen 518055, China}
\affiliation{Center for Quantum Computing, Peng Cheng Laboratory, Shenzhen 518055, China}
\affiliation{Institute for Quantum Computing and Department of Physics and Astronomy,
University of Waterloo, Waterloo N2L 3G1, Ontario, Canada}
\affiliation{State Key Laboratory of Low-Dimensional Quantum Physics and Department of Physics, Tsinghua University, Beijing 100084, China}

\author{Muxin Han}
\thanks{These authors contributed equally to this work.}
\affiliation{Department of Physics, Florida Atlantic University, 777 Glades Road, Boca Raton, FL 33431, USA}
\affiliation{Institut f\"ur Quantengravitation, Universit\"at Erlangen-N\"urnberg, Staudtstr. 7/B2, 91058 Erlangen, Germany}

\author{Dongxue Qu}
\affiliation{Department of Physics, Florida Atlantic University, 777 Glades Road, Boca Raton, FL 33431, USA}

\author{Zichang Huang}
\affiliation{Department of Physics, Florida Atlantic University, 777 Glades Road, Boca Raton, FL 33431, USA}

\author{Guilu Long}
\affiliation{State Key Laboratory of Low-Dimensional Quantum Physics and Department of Physics, Tsinghua University, Beijing 100084, China}

\author{Yidun Wan}
\email{ydwan@fudan.edu.cn}
\affiliation{State Key Laboratory of Surface Physics, Fudan University, Shanghai 200433, China}
\affiliation{Department of Physics and Center for Field Theory and Particle Physics, Fudan University,
Shanghai 200433, China}
\affiliation{Institute for Nanoelectronic devices and Quantum computing, Fudan University, Shanghai 200433,
China}
\affiliation{Shenzhen Institute for Quantum Science and Engineering, and Department of Physics, Southern University of Science and Technology, Shenzhen 518055, China}
\affiliation{Center for Quantum Computing, Peng Cheng Laboratory, Shenzhen 518055, China}
\affiliation{Collaborative Innovation Center of Advanced Microstructures, Nanjing University, Nanjing, 210093, China}

\author{Dawei Lu}
\email{ludw@sustech.edu.cn}
\affiliation{Shenzhen Institute for Quantum Science and Engineering, and Department of Physics, Southern University of Science and Technology, Shenzhen 518055, China}
\affiliation{Center for Quantum Computing, Peng Cheng Laboratory, Shenzhen 518055, China}
\affiliation{Shenzhen Key Laboratory of Quantum Science and Engineering, Shenzhen 518055, China}

\author{Bei Zeng}
\affiliation{Shenzhen Institute for Quantum Science and Engineering, and Department of Physics, Southern University of Science and Technology, Shenzhen 518055, China}
\affiliation{Center for Quantum Computing, Peng Cheng Laboratory, Shenzhen 518055, China}
\affiliation{Institute for Quantum Computing and Department of Physics and Astronomy, University of Waterloo, Waterloo N2L 3G1, Ontario, Canada}
\affiliation{Department of Mathematics and Statistics, University of Guelph, Guelph N1G 2W1, Ontario, Canada}

\author{Raymond Laflamme}
\affiliation{Institute for Quantum Computing and Department of Physics and Astronomy,
University of Waterloo, Waterloo N2L 3G1, Ontario, Canada}
\affiliation{Perimeter Institute for Theoretical Physics, Waterloo N2L 2Y5, Ontario,
Canada}%

\date{\today}
\maketitle
%\textbf{Anti-de Sitter/conformal field theory (\textsc{a}d\textsc{s/cft}) correspondence is one of the most promising realizations of holographic principle towards quantum gravity. The recent
%development of a discrete version of \adscft\ correspondence in terms of tensor networks motivates
%one to simulate and demonstrate \adscft correspondence on quantum simulators. We achieve this
%goal indeed, in this work, on
%a six-qubit nuclear magnetic resonance quantum simulator. We demonstrate experimentally the discrete \adscft correspondence,
%under realistic noises, by measuring the relevant entanglement entropies on the corresponding
%tensor network state. The fidelity of our experimentally prepared tensor network state is 85.0\% via full state tomography and rises to 93.7\% if the signal-decay due to decoherence is taken into account. Our experiment serves as the basic module of simulating more complex
%tensor network states that exhibit \adscft correspondence. As the initial experimental attempt to study \adscft via quantum information processing, our work opens up new avenues exploring quantum gravity phenomena on quantum simulators.}

\textbf{Quantum simulation promises to have wide applications in many fields where problems are hard to model with classical computers. Various quantum devices of different platforms have been built to tackle the problems in, say, quantum chemistry, condensed matter physics, and high-energy physics.  Here, we report an experiment towards  the simulation of quantum gravity by simulating the holographic entanglement entropy. On a six-qubit nuclear magnetic resonance quantum simulator, we demonstrate a key result of Anti-de Sitter/conformal field theory(\adscft) correspondence---the Ryu-Takayanagi formula is demonstrated by measuring the relevant entanglement entropies on the perfect tensor state. The fidelity of our experimentally prepared the six-qubit state is 85.0\% via full state tomography and reaches 93.7\% if the signal-decay due to decoherence is taken into account. Our experiment serves as the basic module of simulating more complex tensor network states that exploring \adscft correspondence. As the initial experimental attempt to study \adscft via quantum information processing, our work opens up new avenues exploring quantum gravity phenomena on quantum simulators.}

%(NMR), since the NMR provides an easily-operable quantum simulation
%platform with long decoherence time

\section{Introduction}
The study of quantum systems requires an exponential amount of resources on
conventional computers due to the exponentially growing dimensionality of Hilbert spaces, which makes it impossible to model even with supercomputers. Quantum simulators, conceived by Feymann in 1982\cite{1982Feynman}, are special purpose devices designed to imitate the behaviours or properties of other less accessible quantum systems\cite{Lloyd1073,2014RMPquantumsimulation}. Over the past few years, proof-of-principle experiments have been realized in simulating quantum phase transitions\cite{2002phase,peng2005quantum}, topological order\cite{2008topoQS,you2010quantum}, molecular energies\cite{2010moleculer,lanyon2010towards}, quantum chaos\cite{2002chaos,howell1999linear}, and so on. However, there exists a significant field -- quantum gravity -- that has never been explored by experimental quantum simulation. Many important ideas such as holographic principle and \adscft correspondence remained unrevealed in experiment. Recent development of a discrete version of \adscft\ correspondence in terms of tensor networks motivates
us to studying \adscft correspondence on quantum simulators. In this work, we make first steps toward to the simulation of quantum gravity on a 6-qubit nuclear magnetic resonance (NMR) quantum processor, where rank-6 perfect tensor that forms the building block of complex tensor networks is realized with high accuracy.

We start from a basic introduction to \adscft\ correspondence. \textsc{a}d\textsc{s/cft} correspondence is one of the most prominent approaches towards a quantum theory of gravity for over two decades \cite{Maldacena:1997re,Aharony:1999ti}. It is the most successful realization of the holographic principle to date, by stating that the quantum gravity theory in the bulk anti-de Sitter spacetime is equivalent to a quantum conformal field theory on the lower-dimensional boundary of the spacetime. The \adscft\ correspondence has recently become a bridge connecting quantum gravity to quantum information theory \cite{Maldacena:2013xja,Brown:2015bva}, which inspires revolutionary ideas of developing quantum gravity using the methods in quantum information and entanglement. %Within this new framework, quantum gravity can be understood via exactly solvable quantum models. For instance, a black hole is related to quantum scrambling in the chaotic systems \cite{Maldacena:2015waa,Hosur:2015ylk,Roberts:2016hpo} and studied in the Sachdev-Ye-Kitaev model \cite{Kitaev,PhysRevLett.70.3339,Maldacena:2016hyu}; the spacetime geometry emerges from entanglement \cite{Ryu:2006bv,VanRaamsdonk:2016exw}, which can be realized by tensor networks and quantum error correcting codes \cite{Almheiri:2014lwa,Pastawski:2015qua,Qi1,Freedman:2016zud}. The holographic duality of \adscft offers a dictionary between the observables of the $d$-dimensional bulk gravity theory and those of the $(d-1)$-dimensional boundary field theory, i.e. properties of the bulk gravity and geometry can be reconstructed or emerged from the boundary field theory (known as the emergent gravity program \cite{VanRaamsdonk:2016exw}), and vice versa. In particular, 
A key result in this perspective is the holographic entanglement entropy characterized by the Ryu-Takayanagi (\scap{rt}) formula, which relates the entanglement entropy of the boundary quantum system to the bulk geometry:
 \be
S_{EE}(\mathcal{A}) = \frac{\Ar_{\textrm{min}}}{4G_N},\label{RT}
\ee
$S_{EE}(\mathcal{A})$ is the entanglement entropy of a $(d-1)$-dimensional boundary region $\mathcal{A}$, while $\Ar_{\textrm{min}}$ is the area of the bulk $(d-2)$-dimensional minimal surface anchored to $\mathcal{A}$ \cite{Ryu:2006bv,Lewkowycz:2013nqa,dche}. $G_N$ is the Newton constant. See Fig. \ref{adscftAll}(a) for a brief illustration.

Recently, a discrete version of \adscft is realized on a type of lattices called tensor networks (\scap{tn}) \cite{Swingle:2009bg,Pastawski:2015qua,Qi1,HanHung}, making it possible to be demonstrated on a quantum simulator device in practice. In general, \scap{tn} states are ways of rewriting a many-body wave function in terms of contractions of tensors, aiming at obtaining the ground states of interacting many-body Hamiltonians in a numerically efficient way. As a key observation related to \adscft, the \scap{tn} state has an emergent bulk dimension built by the layers of tensors, making it an ideal ground for manifesting \adscft in many-body systems. Indeed, the theoretical studies have found that the \scap{tn} made of \emph{perfect tensors} (\scap{pt}) can demonstrate interesting holographic properties. In particular, the entanglement entropy of perfect tensor \scap{tn} gives a discrete realization the above \scap{rt} formula \cite{Pastawski:2015qua}.

In this work, we demonstrate the \scap{rt} formula %and hence \adscft 
on a quantum simulator that simulates a \scap{pt} of rank-$6$. Using a six-qubit quantum register in the nuclear magnetic resonance (\scap{nmr}) system, we create the rank-$6$ \scap{pt} and subsequently measure its holographic entanglement entropy. The experimental results demonstrate the \scap{rt} formula if the decoherence effect is taken into account. As the rank-$6$ \scap{pt} serves as the building block to construct the entire \scap{tn}, our experiment also opens up a new and practical way of studying \adscft and the holographic principle at large.

\section{Results}

\begin{figure*}[!ht]
%\centering
\includegraphics[width= 0.6\columnwidth]{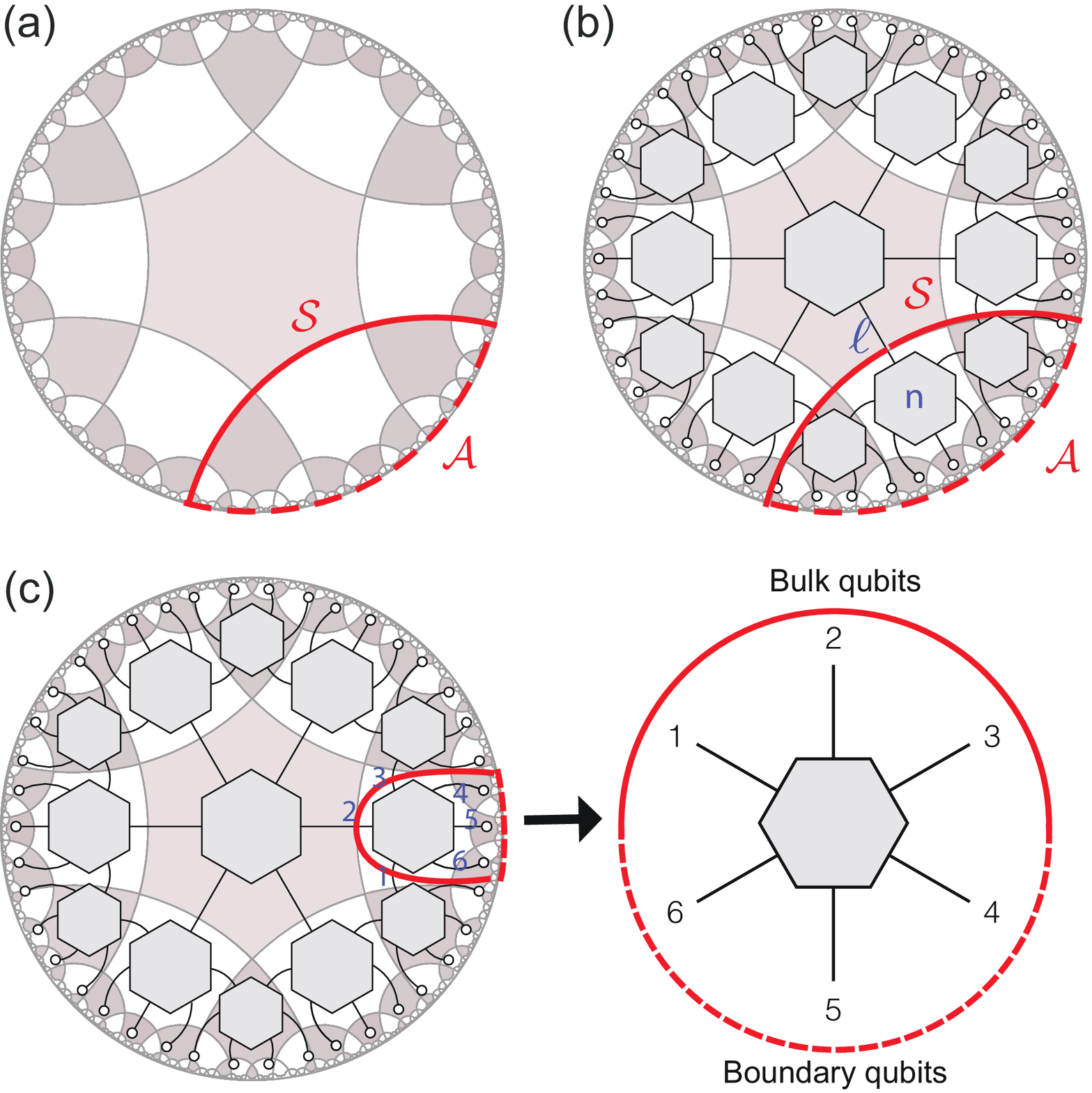}
\caption{(a) A sketch of the \scap{rt} formula. The hexagonal tiling indicates that the disk is a $2$-dimensional
\textsc{a}d\textsc{s} space. The red solid arc in the bulk is the minimal surface (a line in this case) anchored to the two ends of a chosen boundary region $\mathcal{A}$. (b) A discretization of (a) by a
tensor network comprised of rank-$6$ tensors. Each hexagonal node represents a rank-$6$ tensor state $|\psi\rangle\in \ch^{\otimes 6}$, and the collection of all such nodes corresponds to the tensor product of all $|\psi\rangle$'s. Each link $\ell$ represents a maximally entangled state $|\ell\rangle= \left( \ket{00}+\ket{11} \right) / \sqrt{2}$. Connecting one leg of the node to a link corresponds to taking the inner product in $\ch$. The dangling legs are physical qubits in the many-body system. The red dashed arc illustrates the virtual surface $\cs$ anchored to region $\mathcal{A}$, which cuts a minimal number of links. (c)  Rank-$6$ \scap{pt} from the \scap{tn} with the minimal number of cuts equal to three. The six legs represent six qubits. Three qubits are at the boundary and the other three are bulk qubits. This is the model realized in our experiment.}
\label{adscftAll}
\end{figure*}

\emph{Perfect tensors} -- The \scap{tn} that we focus on is shown in Fig. \ref{adscftAll}(b), where each hexagon represents a special six-qubit state $\ket{\psi}$. $\ket{\psi}$ is called a \scap{pt}, if and only if that any three-qubit subsystem out of six is maximally entangled with the rest. It is shown that, for a \scap{tn} made by the \scap{pt}, its entanglement entropy is holographic and gives the discrete \scap{rt} formula on the lattice. Actually, the entanglement entropy of such \scap{tn} equals the \emph{minimal} number of links cut by the virtual surfaces anchored to the boundary, as illustrated in Fig. \ref{adscftAll}(b).

To prove the above statement, we first introduce the form of the rank-$6$ \scap{pt}, which is the building block of the \scap{tn}. Given the single-qubit Hilbert space $\ch\simeq\C^2$, a rank-$6$ \scap{pt} $|\psi\rangle$ is a state in $\ch^{\otimes 6}$, such that for any bipartition of qubits $m+k=6$, the entropy of the reduced density matrix is maximal. Assuming $m \geq k$, and labeling the orthonormal basis in $\ch^{\otimes m}$ and $\ch^{\otimes k}$ by $|\a\rangle$ and $|i\rangle$ respectively, a \scap{pt} $|\psi\rangle=\sum_{\a,i}\psi_{\a i}|\a\rangle\otimes|i\rangle$ satisfies
\be
\sum_{\a}{\psi^\dagger}_{i\a}\psi_{\a j}=\frac{1}{2^k}\delta_{ij}.\label{perfect}
\ee
In other words, the reduced density matrix $\rho^{(k)}$ by tracing out $m$ qubits is an identity matrix, whose entanglement entropy
\be
 S^{(k)}_{EE} = -\tr \left( \rho^{(k)} \log_2 \rho^{(k)} \right)
 \label{SEE}
 \ee
 is simply $k$, the number of remained qubits. In this Letter, we use the superscript $(k)$ to represent the $k$-qubit subsystem.

With the rank-$6$ \scap{pt} (explicit form in appendix A \cite{supple}) in hand, the \scap{tn} state illustrated in Fig. \ref{adscftAll}(b) is constructed as follows. Each internal link $\ell$ represents a two-qubit maximally entangled state $|\ell\rangle= \left( \ket{00}+\ket{11} \right) / \sqrt{2}$, where two qubits associate respectively to the two end points of $\ell$. If we denote by $|\psi(n)\rangle$ the \scap{pt} associated to the hexagon node $n$, the total \scap{tn} state $|\Psi\rangle$ in Fig. \ref{adscftAll}(b) is written as a (partial) inner product form
\be
|\Psi\rangle=\bigotimes_\ell\langle\ell|\bigotimes_n|\psi(n)\rangle.
\ee
The inner product takes place at the end points of each internal link $\ell$, between one qubit in $|\ell\rangle$ and the other in $|\psi(n)\rangle$. The qubits in $|\psi(n)\rangle$ not participating the inner product are boundary qubits corresponding to the dangling legs, and these boundary ones are actually physical qubits, indicating that $|\Psi\rangle$ is a state on the boundary.

We then pick a boundary region $\mathcal{A}$ which collects a subset of the boundary qubits, as shown in Fig. \ref{adscftAll}(b). The reduced density matrix $\rho_\mathcal{A}=\tr_{\bar{\mathcal{A}}}\left( |\Psi\rangle\langle\Psi| \right)$ is computed by tracing out all boundary qubits outside $\mathcal{A}$. Initially, this partial trace boils down to computing the reduced density matrix of individual tensors closest to the boundary. By applying Eq. (\ref{perfect}) and noticing that $|\ell\rangle$ is maximally entangled, the trace computation can be effectively pushed from the boundary into the bulk, meaning that the partial trace on the boundary is now equivalent to computing the reduced density matrix of the \scap{pt} inside the bulk \cite{supple}. Once again, we can apply Eq. \Ref{perfect} and push the trace further inside. This iteration procedure is repeated until the trace reaches $\cs$ in Fig. \ref{adscftAll}(b), where Eq. \Ref{perfect} is not anymore valid, as the number of qubits participating the trace (number of links cut by $\cs$) is less than three for each tensor.

Now we have presented a sketch about how to calculate the entanglement entropy of $\rho_\ca$ via Eq. (\ref{SEE}), and direct readers to appendix B \cite{supple} for a concise proof using the graphical computation of \scap{tn}. Firstly, $\tr\left( \rho_\mathcal{A} \right)$ is found to be equal to the number of qubits on $\cs$, i.e. the same as the number of links cut by $\cs$. Moreover, the product $\rho_\ca^2$, involving the inner product of boundary qubits in $\ca$, gives that $\rho_\ca^2\propto\rho_\ca$. Note that we have ignored all numerical prefactors but they all cancel when calculating $\frac{\tr\rho_\ca^n}{(\tr\rho_\ca)^n}$ in the entanglement entropy. As a result, the Von Neumann entropy gives \cite{supple}
\begin{eqnarray}
S_{EE}(\ca)&=&\lim_{n\to 1} \frac{1}{1-n}\log_2 \frac{\tr\rho_\ca^n}{(\tr\rho_\ca)^n} \nonumber \\
&= &\text{minimal number of cuts by }\cs.
\label{ent}
\end{eqnarray}
The above result is a discrete version of the \scap{rt} formula in Eq. \Ref{RT}. The ``minimal number of cuts'' represents the minimal area $\Ar_{\rm min}$ (in the unit of Planck scale) in the \scap{rt} formula. The bulk surface $\cs$ with minimal area emerges effectively from the entanglement entropy of the \scap{tn} state. Eq. \Ref{ent} demonstrates explicitly that the bulk geometry are created holographically by the entangled qubits of the boundary many-body system.

It is worth emphasizing that, all descriptions about constructing the \scap{tn} originate from the \scap{pt} in Eq. (\ref{perfect}). Therefore, this rank-$6$ \scap{pt} plays the fundamental role in holographic entanglement entropy, and is a key of emerging bulk gravity from \scap{tn} states. If we choose $\cs$ as shown in Fig. \ref{adscftAll}(c) by which the minimal number of cuts is three, a rank-$6$ \scap{pt} is generated where the boundary and bulk qubits are both three. Here, we demonstrate the emergent gravity program in \adscft for the first time in a six-qubit \scap{nmr} quantum simulator, by creating the rank-$6$ \scap{pt} in Fig. \ref{adscftAll}(c) and measuring the relevant entanglement entropies.

\begin{figure*}[!ht]
\centering
\includegraphics[width= 0.9 \columnwidth]{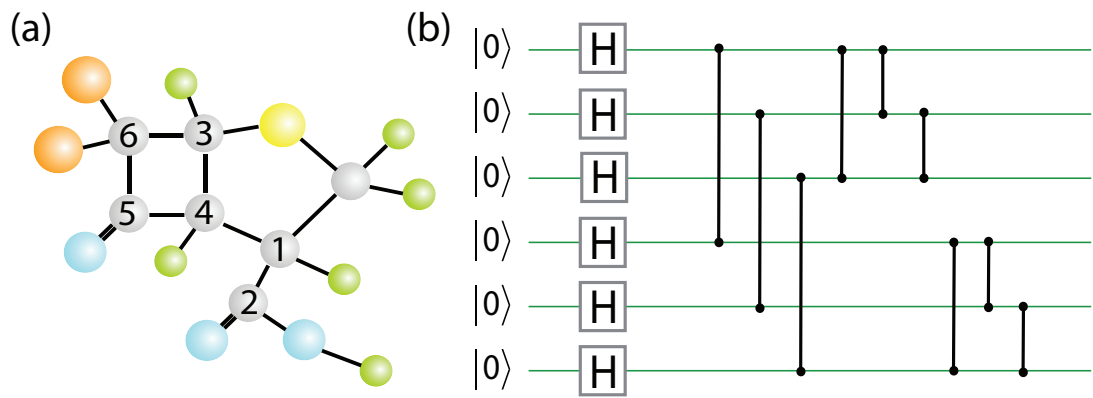}
\caption{ (a) Molecular structure of the $^{13}$C-labeled six-qubit quantum processor. The six qubits of the rank-$6$ \scap{pt} are mapped to \emph{1} to \emph{6}, respectively. (b) Quantum circuit that evolves the system from $\ket{0}^{\otimes 6}$ to the \scap{pt}, constructed by several Hadamard gates (blocks) and controlled-$Z$ operations (lines connecting two dots).}
\label{exp_fig1}
\end{figure*}

\emph{Experiment implementation of a rank-$6$ perfect tensor} -- The six qubits in the \scap{nmr} quantum register are denoted by the spin-1/2 $^{13}$C nuclear spins, labeled as \emph{1} to \emph{6} as shown in Fig. \ref{exp_fig1}(a), in $^{13}$C-labeled Dichloro-cyclobutanone dissolved in d$_6$-acetone. All experiments were carried out on a Bruker \scap{drx} 700 \scap{mh}z spectrometer at room temperature. The internal Hamiltonian of this system is
\begin{align}\label{Hamiltonian}
\mathcal{H}_{int}=\sum\limits_{j=1}^6 {\pi \nu _j } \sigma_z^j  + \sum\limits_{j < k,=1}^6 \frac{\pi}{2} J_{jk} \sigma_z^j \sigma_z^k,
\end{align}
where $\nu_j$ is the resonance frequency of the \emph{j}th spin and $\emph{J}_{jk}$ is the $J$-coupling strength between spins \emph{j} and \emph{k}. All parameters including the relaxation times for each spin are listed in appendix C \cite{supple}. To control system dynamics, we have external control pulses with four adjustable parameters: the amplitude, frequency, phase, and duration, based on which arbitrary single-qubit rotations can be realized with simulated fidelities over 99.5\% \cite{supple}.

A rank-$6$ \scap{pt} can be created from $\ket{0}^{\otimes n}$ through the circuit as illustrated in Fig. \ref{exp_fig1}(b), which involves only Hadamard gates and controlled-$Z$ gates. Experimentally, this requires an
%The ordinary way of preparing entangled states in quantum computing is first
initialization of the system onto $\ket{0}^{\otimes n}$.
% -- which is usually made up of Hadamard gates and controlled-\scap{not} gates -- to entangle qubits. For instance, the rank-$6$ \scap{pt} can be created via the circuit in Fig. \ref{exp_fig1}(b).
However, initializing an \scap{nmr} processor to $\ket{0}^{\otimes n}$ is based upon the pseudo-pure state technique, which leads to an exponential signal attenuation. Here, we adopt a temporal averaging approach that enables the \scap{pt} preparation directly from the thermal equilibrium of \scap{nmr}, while skipping the intermediate pseudo-pure state stage to avoid the above problem, as shown in appendix C \cite{supple}).

After the creation, we conducted $k$-qubit ($1\leq k \leq 5$) quantum state tomography in the corresponding subspace of the whole system, respectively. For simplicity, the cutting of the links was chosen to be continuous in experiment, i.e. in a cyclic manner. It means that six state tomographies for any given $k$ are performed, e.g. when $k=2$, we reconstructed $\rho^{(2)}_{12}, \rho^{(2)}_{23}, \cdots, \rho^{(2)}_{61}$.

\begin{figure*}[!ht]
\centering
\includegraphics[width= 0.95\columnwidth]{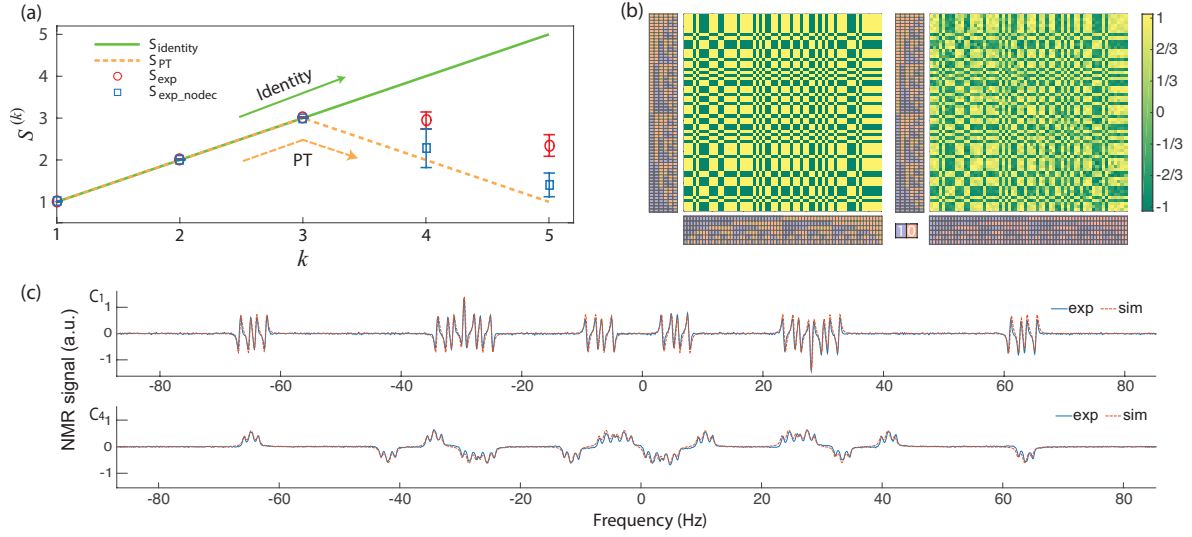}
\caption{(a) Entanglement entropy $S^{(k)}$ of the $k$-qubit subsystem of the rank-$6$ \scap{pt}. In theory, $S^{(k)} = \text{min}\{k,6-k\}$ as shown by the orange dashed line. Experimental results are represented by the red circles, where $S^{(4)}$ and $S^{(5)}$ do not fit very well. If the signal's decay due to decoherence is taken into account, the experimental results are rescaled to the blue squares, which fit much better. As a upper-bound reference, the maximal entropy of a $k$-qubit subsystem is also plotted (green dotted line) by assuming a six-qubit identity. (b) Density matrices of the theoretical rank-$6$ \scap{pt} $\rho_{pt}$ (left) and the experimentally reconstructed state $\rho_{e}$ (right) on a two-dimensional plane. The rows and columns are labeled by the six-qubit computational basis from $\ket{0}^{\otimes 6}$ to $\ket{1}^{\otimes 6}$, respectively. (c) Direct observation of $\rho_{e}$ in the \scap{nmr} spectra (red), with probe qubits C$_1$ (top) and C$_4$ (bottom), respectively. The simulated spectra of the \scap{pt} are also shown in blue. For a better visualization, experimental signals are rescaled by 1.25 times to neutralize the decoherence error.}
\label{exp_fig2}
\end{figure*}

Subsequently, Von Neumann entropies of these $k$-qubit subsystems were calculated by Eq. (\ref{SEE}).
In theory, each $S^{(k)}$ equals to the minimal number of cuts according to Eq. (\ref{ent}) when $k\leq 3$. Combined with the fact that $S^{(k)} = S^{(6-k)}$ for a six-qubit pure state, we have $S^{(k)} = \text{min}\{k,6-k\}$ for the theoretical \scap{pt}, as shown by the orange dashed line in Fig. \ref{exp_fig2}(a). In experiment however, inevitable errors lead to imperfection and hence impurity in the truly prepared state, so we cannot just measure $k\leq 3$ cases to deduce other $k$'s. Therefore, we measured and compare the experimental $S^{(k)}$ for each $1\leq k \leq 5$ (red circles) with their theoretical predictions in Fig. \ref{exp_fig2}(a). For each $k$, the mean and error bar of the experimental $S^{(k)}$ value are calculated from the six \emph{cyclic} tomographic results. When $k\leq3$, the measured entanglement entropies match extremely well with the theory; when $k>3$, there are notable discrepancies between theory and experiment, which should be primarily attributed to decoherence errors, as discussed in the following.

The pulse sequence that creates the \scap{pt} is around 60 ms; this is not a negligible length compared to the $T_2^*$ time ($\sim400$ ms) of the molecule, meaning that decoherence will induce substantial errors during experiments(see Method). As $T_2^*$ relaxation is the dominating factor, the off-diagonal terms in the \scap{pt} density matrix are mainly affected. To estimate this imperfection, we performed full state tomography \cite{PhysRevA.69.052302} on the prepared state and got $\rho_{e}$. The real part of $\rho_{e}$ is depicted in the right panel of Fig. \ref{exp_fig2}(b), by projecting each element onto a two-dimensional plane. As a comparison, the figure of the theoretical \scap{pt} $\rho_{pt} = \ket{\psi}\bra{\psi}$ is placed in the left panel of Fig. \ref{exp_fig2}(b). In fact, the diagonal elements of $\rho_{e}$ are almost the same as that of $\rho_{pt}$, but the off-diagonal are lower due to the $T_2^*$ errors. The state fidelity between $\rho_{e}$ and $\rho_{pt}$, defined as
\be
F\left(\rho_{pt}, \rho_{e} \right) =\tr \left[ \sqrt{\sqrt{\rho_{pt}} \rho_{e} \sqrt{\rho_{pt}}} \right],
\label{fidelity}
\ee
is about 85.0\%. Direct observations of $\rho_{e}$ in terms of \scap{nmr} spectra are also shown in Fig. \ref{exp_fig2}(c), where experimental and simulated spectra highly match if the experimental signal is rescaled by 1.25 times to compensate for the decoherence effect.

Although the reconstructed state $\rho_{e}$ is prone to the decoherence errors, the entanglement entropies for the cases $k\leq 3$ in Fig. \ref{exp_fig2}(a) are still in excellent accordance with the theory. The reason is, when we trace out three or more qubits, the reduced density matrix is predicted to be identity according to Eq. (\ref{perfect}), so the measured $k\leq3$ reduced density matrices are almost irrelevant to the imperfection of the off-diagonal elements in $\rho_{e}$. However, when $k>3$, the reduced density matrix is no longer the identity, meaning that the imperfect off-diagonal terms in $\rho_{e}$ start to be responsible for calculating $S^{(k)}$. As a result, in Fig. \ref{exp_fig2}(a) we have $S^{(4)} = 2.91\pm 0.20$ and $S^{(5)} = 2.32\pm 0.25$ (red circles) respectively, which are quite distant from the theoretical curve. After numerically simulating and compensating for the decoherence errors \cite{vandersypen2001experimental,PhysRevLett.114.140505} during the \scap{pt} creation, we found that the two entanglement entropies $S^{(4)}$ and $S^{(5)}$ approach much closer to the theory, which are now $2.27\pm 0.46$ and $1.37\pm 0.28$ (blue squares), respectively. We also calculated the current fidelity between the rescaled experimental state and $\rho_{pt}$ via Eq. (\ref{fidelity}), and found it improved to 93.7\%, which is 8.7\% greater than that of $\rho_{e}$.

\section{Discussion}

\scap{Rt} formula, or explicitly the \scap{tn} built by the rank-$6$ \scap{pt} in Fig. \ref{adscftAll}(b), tells us how to deduce the bulk geometry using the entanglement on the boundary. The implicit condition here is that the global \scap{tn} state is \emph{pure}. Otherwise, the information on the boundary cannot uniquely (up to local unitaries) determine the bulk geometry, e.g. it cannot specify whether the \scap{tn} state is the maximally \emph{mixed} identity or \scap{pt} since both give the same entanglement entropies on the boundary (meaning $k\leq 3$) as shown in Fig. \ref{exp_fig2}(a). In experiments, however, under realistic noises,
it is difficult to guarantee
the purity of the truly created states  because experimental procedures inevitably involve errors---in particular the decoherence that  render the \scap{tn} states \emph{mixed}.  In our experiment
of a $6$-qubit \scap{pt}---a build block of a complex \scap{tn}, we have achieved $85\%$
fidelity, which is already state-of-the-art; however, there is yet some non-negligible decoherence
due to the $T_2^*$ errors. Therefore, our results successfully test the \scap{rt} formula up to
the decoherence.

The simulation of the holographic entanglement entropy can be generalized to \scap{tn}s with multiple perfect tensors. In the Section E of the supplemental material, we demonstrate a simulation of the holographic entanglement entropy on a \scap{tn} with 7 tensors. The key to performing the simulation is that measuring the R\'enyi entropies of the \scap{tn}s can be reduced to measuring the reduced density matrices of $\rho_e$, and their multiplications and traces, while $\rho_e$ is simulated experimentally. The result of the simulation demonstrates agreement with the RT formula, up to the experimental noise in $\rho_e$. The simulation can be generalized to other \scap{tn}s.

In conclusion, our work is an endeavour to demonstrate on a quantum simulator the \scap{rt} formula (the discrete \scap{pt} version) in the \adscft correspondence. We utilize a temporal average technique to create the rank-$6$ \scap{pt} and perform full state tomography to reconstruct the experimental state. This is also the largest full state characterization in an \scap{nmr} system to date. Although the imperfection of the created state due to decoherence errors makes the holographic entanglement entropy not exactly agree with the theoretical prediction, we simulate and compensate for such type of errors under the realistic experimental environment, and demonstrate the accordance between theory and experiment thereafter. As the first step towards exploring \adscft correspondence using a quantum simulator, our work provides valid experimental demonstrations about studying quantum gravity in the presence of realistic noises.

\section{method}
\emph{Decoherence simulation} ---To numerically simulate the decoherence effect in our six-qubit system, we made the following assumptions: the environment is Markovian; only the $T_2^*$ dephasing mechanism is taken into account since $T_1$ effect is negligible in our circuit; the dephasing noise is independent between all qubits; the dissipator and the total Hamiltonian commute in each pulse slice as the $\Delta t = 10$ $\mu$s is small. With these assumptions, we simplified and solved the master equation in two steps for each $\Delta t$: evolve the system by the propagator calculated by the internal and control pulse Hamiltonian, and subsequently apply the dephasing factors according to the coherent orders for $\Delta t$ which is an exponential decay of the off-diagonal elements in the density matrix. For each experiment of the 64 runs, we simulated the above process and obtained the signal's decay due to decoherence. From the experimental result, we then compensated for this decay, and a new state in which the decoherence effect was taken into account was thus achieved. The fidelity now between the  rescaled experimental state and the theoretical PT is boosted to 93.7\%.

\bigskip\noindent
\textbf{Data availability}
The data sets generated during and/or analyzed during the current study are available from the corresponding author on reasonable request.

\bigskip\noindent
\textbf{Acknowledgements}
We acknowledge Jianxin Chen, Markus Grassl, Cheng Guo, Ling-Yan Hung, Zhengfeng Ji, Hengyan Wang, and Nengkun Yu for discussions, and anonymous referees for helpful comments. This research was supported by CIFAR, NSERC and Industry of Canada. K.L. and G.L. acknowledge National Natural Science Foundation of China under Grants No. 11774197 and No. 2017YFA0303700.
MH acknowledges support from the US National Science Foundation through grant PHY-1602867, and Start-up Grant at Florida Atlantic University, USA.
YW thanks the hospitality of IQC and
PI during his visit, where this work was partially conducted. YW is partially supported by the Shanghai Pujiang Program grant No. 17PJ1400700 and the NSF grant No. 11875109. D. L. are supported by the  National Natural Science Foundation of China (Grants  No. 11605005, No. 11875159  and No. U1801661),Science, Technology and Innovation Commission of Shenzhen Municipality (Grants No. ZDSYS20170303165926217 and No. JCYJ20170412152620376),  Guangdong Innovative and Entrepreneurial Research Team Program (Grant No. 2016ZT06D348)

\bigskip\noindent
\textbf{Competing Interests}
The authors declare that they have no competing interests.

\bigskip\noindent
\textbf{Author Contributions}
D.L., Y.W., M.H. and R.L. conceived the experiments. K.L. and D.L. performed the experiment and analyzed the data.  M.H. and G.L. provided theoretical support. D.Q. and Z.H. completed the numerical simulation. G.L., B.Z. and R.L. supervised the project. D.L., K.L.,Y.W. and M.H. wrote the manuscript with feedback from all authors. The authors declare that they have no competing financial interests. Correspondence and requests for materials
should be addressed to Y.W.(ydwan@fudan.edu.cn), D.L. (ludw@sustech.edu.cn).
K.L.and M.H. contributed equally to this work.

%\bibliographystyle{naturemag}
%\bibliography{holo_new}% Produces the bibliography via BibTeX.

\end{document}